\documentclass[aps,prb]{revtex4}
\usepackage{graphicx,amsmath,amssymb,bm}

\begin{document}

\title{Topological Structure and an Accurate Wave Function for the Enigmatic 5/2 Fractional Quantum Hall State}	
\author{Sutirtha Mukherjee$^{1,2}$ and Sudhansu S. Mandal$^{1,3,*}$ }
\affiliation{$^1$Department of Theoretical Physics, Indian Association for the Cultivation of Science, Jadavpur, Kolkata 700032, India\\ 
$^2$Korea Institute for Advanced Study, Quantum Universe Center, Seoul 02455, Korea\\	
$^3$Department of Physics and Centre for Theoretical Studies, Indian Institute of Technology, Kharagpur, West Bengal 721302, India	}	

\date{\today}
\maketitle

{\bf 
	 Possibility of relevance of both Moore-Read Pfaffian \cite{Moore91} wave function and its particle-hole conjugate \cite{Levin07,Lee07} partner for the quantized even-denominator  filling factor $5/2$, because they are degenerate for any two-body interaction, have raised confusions about the origin and universality class of the state, and thus other exotic combination of these have also been proposed \cite{Exhotic-1,Exhotic-2}. While the experiments are yet to settle the topological character of the state, on the theoretical front, understanding of even the exact Coulomb ground state is lacking as its overlap \cite{Morf98} with Pfaffian wave function is not satisfactorily high. Here, we comprehensively show the topological structure--order of zeros felt by an electron at the positions of other electrons 
--for the ``Pfaffian-shift" \cite{Wilczek91} in a spherical geometry. A set of linearly independent antisymmetric functions that are constructed from the possible graphs preserving this topological structure provides complete basis for determining an accurate ground state wave function for any interaction, in particular the Coulomb interaction. One of the graphs describes $Z_2$ para-fermion \cite{Read99} equivalently the Pfaffian wave function, but some of the other graphs also contribute to form the  exact ground state for the Coulomb interaction. We further show 
that our wave function for the Coulomb interaction supports clustering up to half of the composite bosons \cite{MG87} signifying a strong-coupling regime.}
	
While the half filling of the lowest Landau level is an unquantized \cite{HLR93} Hall state, the quantized  state has been observed for the same in the second Landau level when the lowest Landau level is completely filled for both up and down spins. Whereas the former state is ascribed to the Fermi sea \cite{HLR93} of composite fermions \cite{Jain89}, the latter with filling factor $\nu =5/2$ has been tagged with the proposed Moore-Read Pfaffian (Pf) \cite{Moore91} state which corresponds to the chiral $p$-wave Cooper pairing \cite{RG00} of composite fermions. The numerical exact diagonalization studies show that the 5/2 state is an incompressible spin-polarized state \cite{Morf98} which has been experimentally verified \cite{Tiemann12} and has a considerable overlap with the Pf wave function.
 The numerical evidence \cite{Scarola00} of effective attractive interaction between composite fermions in the second Landau level is in favour of pairing. The proposed \cite{Moore91,RG00} quasiparticle charge $e/4$ of the Pf state has some experimental support \cite{Dolev08,Radu08,Yacoby11}, although the unquestionable observation of proposed non-abelian braiding  \cite{Moore91,Wilczek91} of quasiparticles is still elusive \cite{Willett13,Baer14}.

The huge confusion arises with the revealing fact that the particle-hole conjugate of the Pf state, namely the anti-Pfaffian (APf) state \cite{Levin07,Lee07} is degenerate for Coulomb interaction and one of these paired state may arise due to spontaneous breaking of particle-hole symmetry. Both the pairing states describe non-abelian braiding of quasiparticles \cite{Moore91,Wilczek91,Wilczek96}, and have edge channel carrying neutral Majorana mode \cite{Wen91} with basic difference is that they support opposite streams of Majorana modes.
Although mixing of Landau levels has been shown to be the possible cause of the breakdown of particle-hole symmetry \cite{Wang09,Rezayi11}, a number of numerical studies produce contrasting results: While some are in favour \cite{Morf98,Wan06,Wojs10,Storni10,Peterson15} of the Pf state, some others indicate APf may  be the relevant state \cite{Rezayi11,Zalatel15}, and even the proposal \cite{Peterson08B,Exhotic-2} of the superposition of Pf and APf wave function has been made. While this issue has neither been settled by theory nor by experiment, we lack satisfactory understanding even at the basic level, {\it i.e.}, about the nature of Coulomb ground state for the Pfaffian flux.
In a spherical geometry, the Pf and APf wave functions become exact zero-energy ground state for a three-body pseudo-potential \cite{Wilczek91} at different number of flux quanta for the same number of electrons; they respectively are $N_\phi = 2N -3$ and $2N+1$ for $N$ electrons. Although this difference defies us for direct comparison between these states,  an accurate form of the ground state wave function at each of these fluxes   
is necessarily required for understanding true nature of the Coulomb ground state at $\nu = 5/2$. This is more so because the overlap between the exact Coulomb ground state and the MRP wave function is not satisfactorily \cite{Morf98,Book} high. In this letter, we take up the case for the Pfaffian flux-shift \cite{Wilczek91}.

Based on the flux-particle relation $N_\phi = 2N-3$, we here propose topological structure, {\it i.e.}, the order of zeros (refraining from taking permutation of the particles) felt by an electron at the positions of other electrons, of 5/2 state, no matter what the underlying interaction is, so long as it produces an incompressible state.
We note that barring the Laughlin wave function, no other wave function has homogeneous topological structure, {\it i.e.}, zeros of a particle are equally distributed at the positions of other particles.
The number of zeros felt by two electrons at each others' position is however same.
 Exploiting this principle, we draw topologically distinct graphs by preserving the topological structure. Each of these graphs produces specific function $f(z_i-z_j)$, where $z_i-z_j = u_iv_j-u_jv_i$ with $u_j = \cos (\theta_j/2)e^{i\phi_j/2},\, v_j =\sin (\theta_j/2) e^{-i\phi_j/2}$ being the spinor coordinates on the surface of a sphere with $0 \leq \theta_j \leq \pi$ and $0 \leq \phi_j \leq 2\pi$. For every system of even number of electrons, we identify a graph which exactly produces Read-Rezayi $Z_2$ para-fermion (Z2P) \cite{Read99} wave function that is equivalent to the Pf wave function. 
 Certain number of other graphs also produce linearly independent functions. The number of linearly independent functions is precisely the Hilbert space dimension in zero-angular momentum sector; each of these functions corresponds to zero angular momentum, by construction. 
 A linear combination of all these functions will form the ground state for a specific interaction. In particular, we numerically show that one such linear combination in a given system of electrons form an accurate Coulomb ground state up to $N=10$.     
Each of the functions produced individually by the connected graphs is non-vanishing even when three or more composite bosons (electrons with one flux quantum attached to each of them) \cite{MG87} come close together, in contrary to the Z2P  wave function \cite{Read99} which vanishes when more than two composite bosons occupy same position. We identify graphs that generate non-zero symmetric functions even when up to half of the composite bosons coalesce.
Since thermodynamically large number of composite bosons can coalesce at a point, the Coulomb interaction between electrons provides strong-coupling regime of composite bosons in 5/2 state.

\begin{figure}[htb]
 	\includegraphics[width=8.0cm]{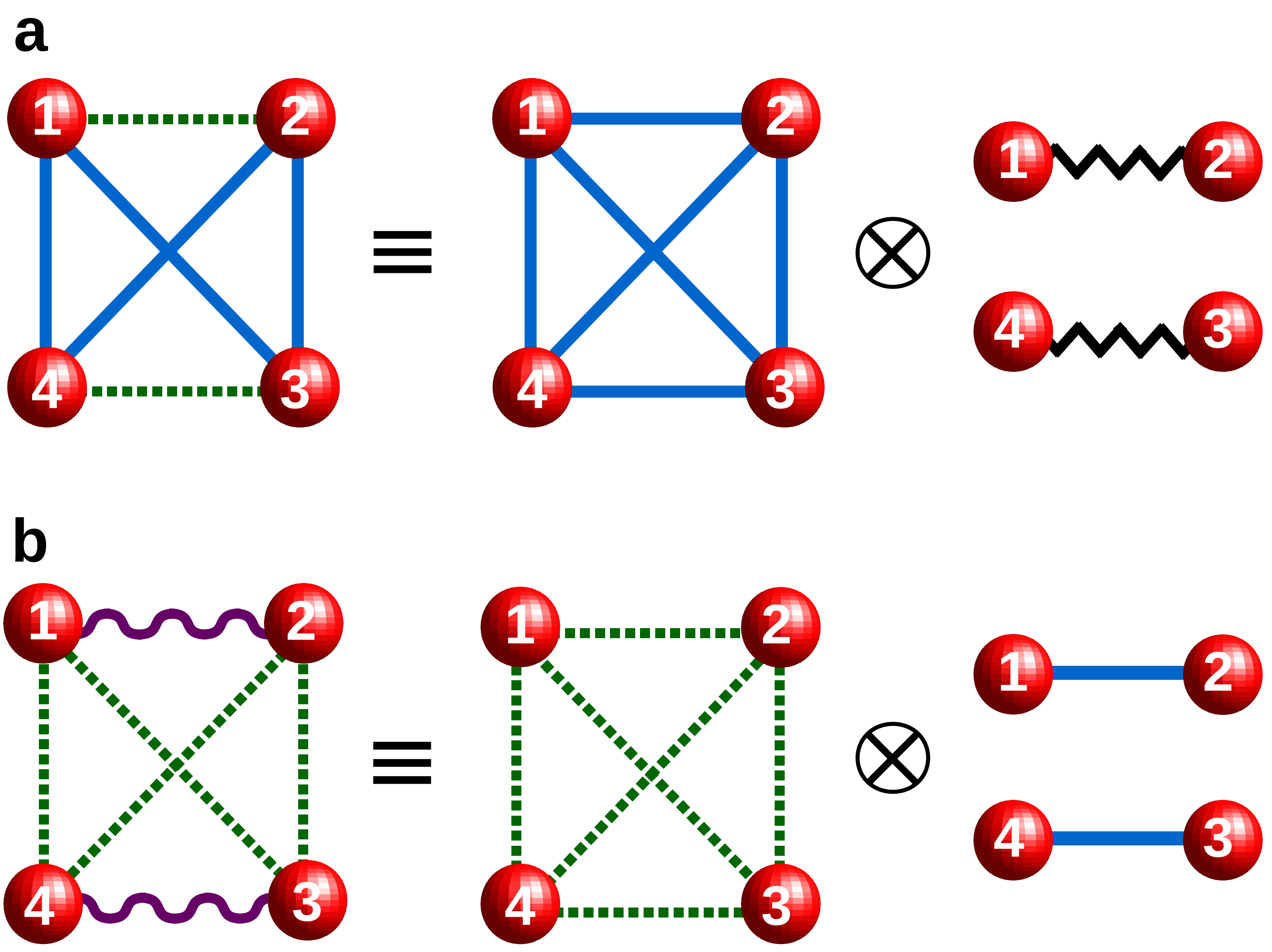}
 	\caption{{\bf Graphs for N=4}. Filled circles with the associated numbers represent different electrons at the vertices. 
 	Links between two electrons represented by dashed (green), solid (blue), sinusoidal (magenta) and wiggly (black) lines describing mutually felt order of
 	zeros at each others' position, namely, 1, 2, 3, and $-1$ respectively. (Readers should not be confused with negative order of zeros as the overall order is always positive). {\bf a} The left graph represents the set ${\cal S}^{(4)} = [1,2,2]$ which can be described by the product of two graphs: The central graph representing the set ${\cal S}^{(4)}_{J,2}=[2,2,2]$ produces Jastrow form $J^2 = \prod_{i<j}^4z_{ij}^2$ and the right-most graph representing the set ${\cal S}^{(4)}_{Pf} = [(-1),0,0]$ produces Pfaffian, ${\cal A}(z_{12}^{-1}z_{34}^{-1})$ where ${\cal A}$ denotes anti-symmetrization when all permutations between the electrons are taken into account. {\bf b } The left diagram represents the set ${\cal S}^{(4)} = [1,1,3]$ which can be described by the product of two graphs: The central graph representing the set ${\cal S}^{(4)}_{J,1}=[1,1,1]$ produces Jastrow form $J = \prod_{i<j}^4z_{ij}$ and the right-most graph representing the set ${\cal S}^{(4)}_{Z2P} = [0,0,2]$ produces ${\cal B} (z_{12}^{2}z_{34}^2)$ where ${\cal B}$ denotes symmetrization when all permutations between the electrons are taken into account.}
 	\label{Fig1}
 \end{figure}

For the smallest system with $N=2$, $N_\phi = 1$ and thus the set of order of zeros (SOZ) is ${\cal S}^{(2)}=[1]$ where superscript denotes the number of electrons. In the second step of recursion, $N=4$ and
$N_\phi=5$, the increase of flux $\Delta N_\phi =4$ for the increase of $N$ by 2. Therefore the additional four zeros can  be accommodated between two additional electrons in two ways which lead to obtain ${\cal S}^{(4)} = [1,2,2]$ or $[1,1,3]$. Both these sets are digrammatically represented by graphs in Fig.~\ref{Fig1}. Each of these is recast by the direct product of two graphs representing two sets: (i) Former is decomposed into two sets ${\cal S}^{(4)}_{J,2} = [2,2,2]$ and ${\cal S}^{(4)}_{Pf} =[(-1),0,0]$; (ii) decomposed two sets for the latter are ${\cal S}^{(4)}_{J,1} = [1,1,1]$ and ${\cal S}^{(4)}_{Z2P}=[0,0,2]$. (The negative order of zero implies the order of singularity which can however be removed by reinstating the decoupled set ${\cal S}_{J,2}^{(4)}$.) While ${\cal S}_{J,n}^{(N)}$ reproduces the Jastrow form of the wave function $J^n = \prod_{i<j}z_{ij}^n$, ${\cal S}_{Pf}^{(4)}$ and ${\cal S}_{Z2P}^{(4)}$ generate the terms $z_{12}^{-1}z_{34}^{-1}$ and $z_{12}^2z_{34}^2$ respectively.
 Therefore the readily obtained respective  functions by taking all permutations of the electrons in the graphs are precisely the
Pf, $\Psi_{Pf}^{(4)}$ and Z2P, $\Psi_{Z2P}^{(4)}$  wave functions respectively for $4$ electrons. These two wave functions are identical up to a normalization factor. Since only one graph (Fig.\ref{Fig1}) is possible for ${\cal S}_{Pf}^{(4)}$ or ${\cal S}_{Z2P}^{(4)}$,  $\Psi_{Pf}^{(4)}$ or $\Psi_{Z2P}^{(4)}$ should be exact ground state wave function, no matter what the interaction potential is. Indeed, the exact ground state wave function for the Coulomb interaction with four electrons has 100\% overlap  with $\Psi_{Pf}^{(4)}$ or $\Psi_{Z2P}^{(4)}$. We note that the Hilbert-space dimension at zero angular momentum is only one.

\begin{figure}[htb]
 	\includegraphics[width=8.0cm]{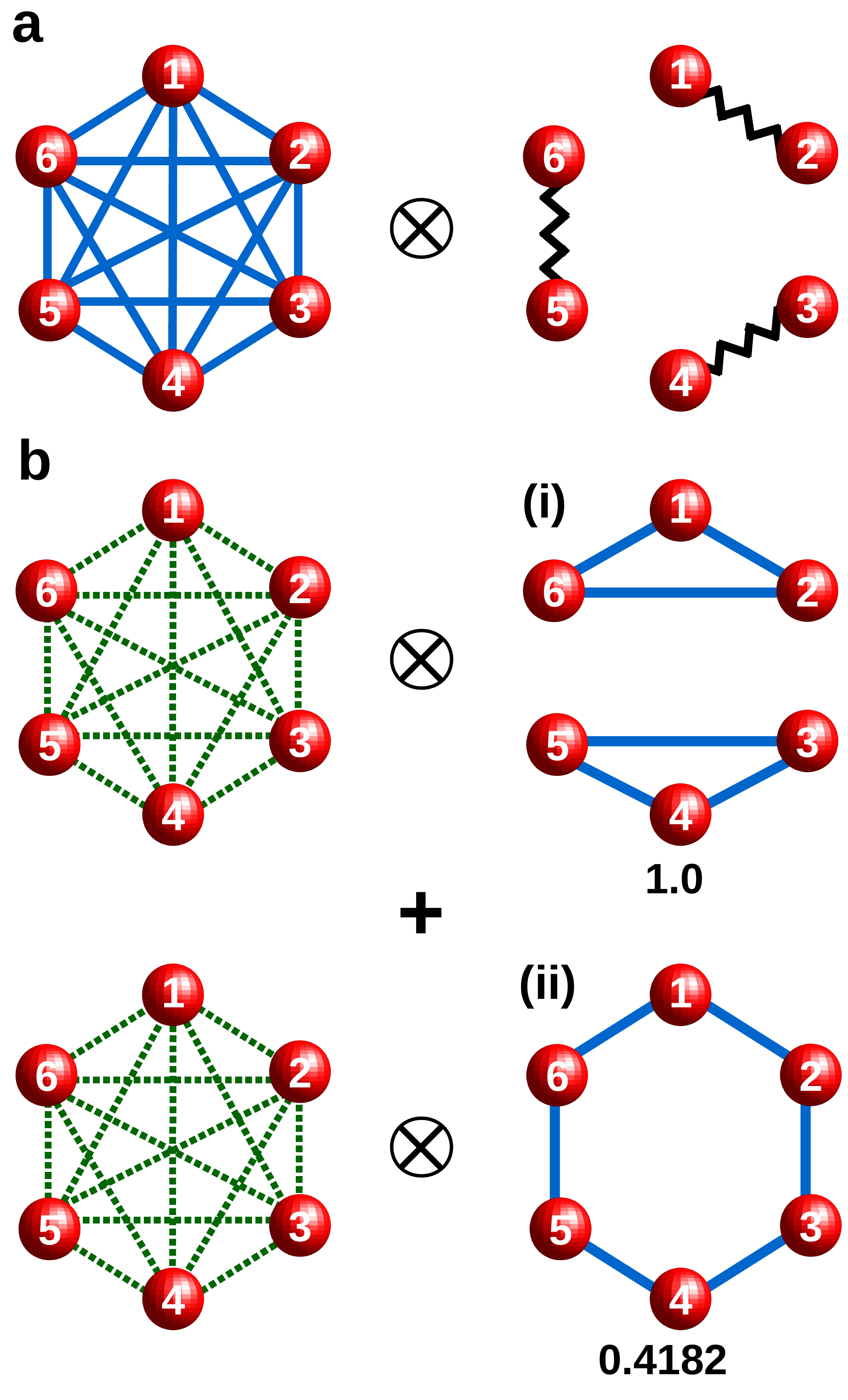}
 	\caption{{\bf Graphs for N=6.} All symbols in the graph are as described in Fig.\ref{Fig1} {\bf a} The product of two graphs: The left graph representing the set ${\cal S}^{(6)}_{J,2}=[2,2,2,2,2]$ produces Jatrow term $J^2$ for 6 electrons and right graph representing the set ${\cal S}^{(6)}_{Pf}=[(-1),0,0,0,0] $ produces the Pfaffian, ${\cal A}(z_{12}^{-1}z_{34}^{-1})$, when all the permutation between electrons are considered. {\cal b} The left graphs of both top and bottom panels representing the set ${\cal S}^{(6)}_{J,1}=[1,1,1,1,1]$ produces Jastrow term $J$ for 6 electrons. The right graphs of both top and bottom panels representing the single set ${\cal S}^{(6)}_{Z2P} = [0,0,0,2,2]$ produces two topologically distinct graphs contributing different symmetric functions upon taking permutations of all electrons. The disconnected graph corresponds to the $Z_2$ para-fermion wave function. The most general wave function is the linear combination between the contributions of both the graphs. The associated numbers represent relative weight factors for the linear combination that provide wave function closest to the exact Coulomb ground state, { \it i.e.}, with overlap 0.99999(0).   }
 	\label{Fig2}
 \end{figure}

A system of six electrons will accommodate four more number of flux quanta than it is for four electrons. Therefore, these extra four zeros increase the elements in the six-particle set as ${\cal S} ^{(6)} = [1,2,2,2,2]$ or $[1,1,1,3,3]$. As discussed before, we then decompose the former set to ${\cal S}_{J,2}^{(6)}=[2,2,2,2,2]$ and ${\cal S}_{Pf}^{(6)}=[(-1),0,0,0,0]$ and the latter set to ${\cal S}_{J,1}^{(6)}=[1,1,1,1,1]$ and $[0,0,0,2,2]$. We can construct only one graph for ${\cal S}_{Pf}^{(6)}$ and hence the corresponding wave function for ${\cal S} ^{(6)}$ is exactly the Pf wave function. On the other hand, two topologically distinct graphs can be formed using ${\cal S}_{Z2P}^{(6)}$: One corresponds to a disconnected graph having two parts, each of which contains three electrons, and the other interestingly describes entanglement between these two parts by breaking one link (bond) in each part and reforming inter-part links. The former graph describes Z2P state for six electrons and the corresponding wave function is identical with the Pf wave function. In general, the wave function derived from the disconnected graph of ${\cal S}_{Z2P}^{(N)}$ is identical with the wave function derivable from the only graph corresponding to ${\cal S}_{Pf}^{(N)}$, we thus ignore ${\cal S}_{Pf}^{(N)}$ in what follows. As studied before, the wave function $\Psi_{Z2P}^{(6)}$ is the exact ground state for a model 3-body pseudopotential, but it has overlap less than $82\%$ only with the ground state wave function for the realistic Coulomb interaction. A linear combination of this wave function with the wave function obtainable from the connected graph (Fig.~\ref{Fig2}) enable us to obtain more than $99.999\%$ overlap (Table-\ref{Table1}) with the exact Coulomb ground state.

\begin{figure}[htb]
 	\includegraphics[width=16.0cm]{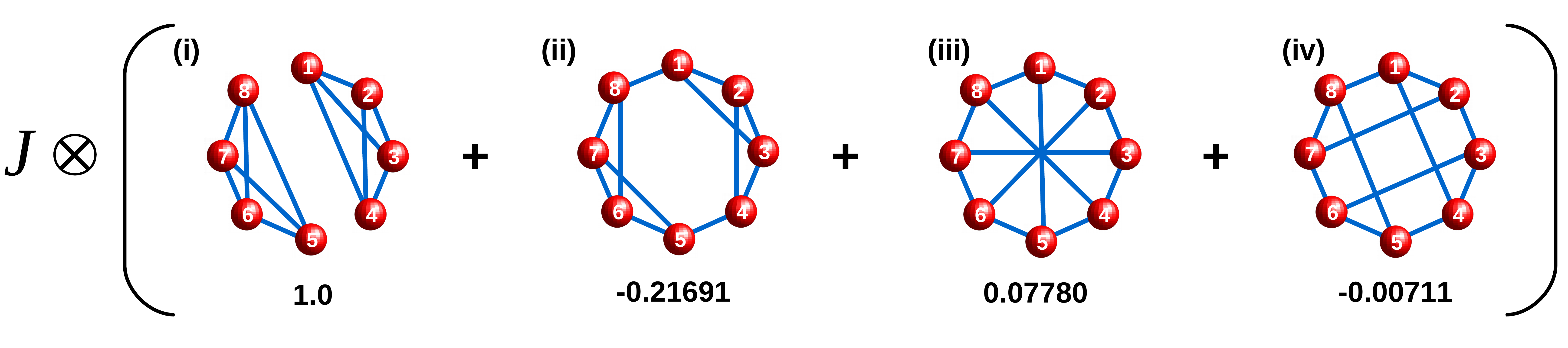}
 	\caption{{\bf Graphs for N=8.} All symbols in the diagram are as described in Fig.\ref{Fig1}. 
 	$J$ represents Jastrow factor for 8 electrons.  Four graphs for the set ${\cal S}_{Z2P}^{(8)}=[0,0,0,0,2,2,2]$ producing linearly independent symmetric functions. The relative weight factors mentioned at the bottom of each graph denote the corresponding coefficients of the symmetric functions that provide 0.99999(0) overlap with the exact ground state wave function for the Coulomb interaction.
 		   }
 	\label{Fig3}
 \end{figure}

\begin{figure}[htb]
 	\includegraphics[width=16.0cm]{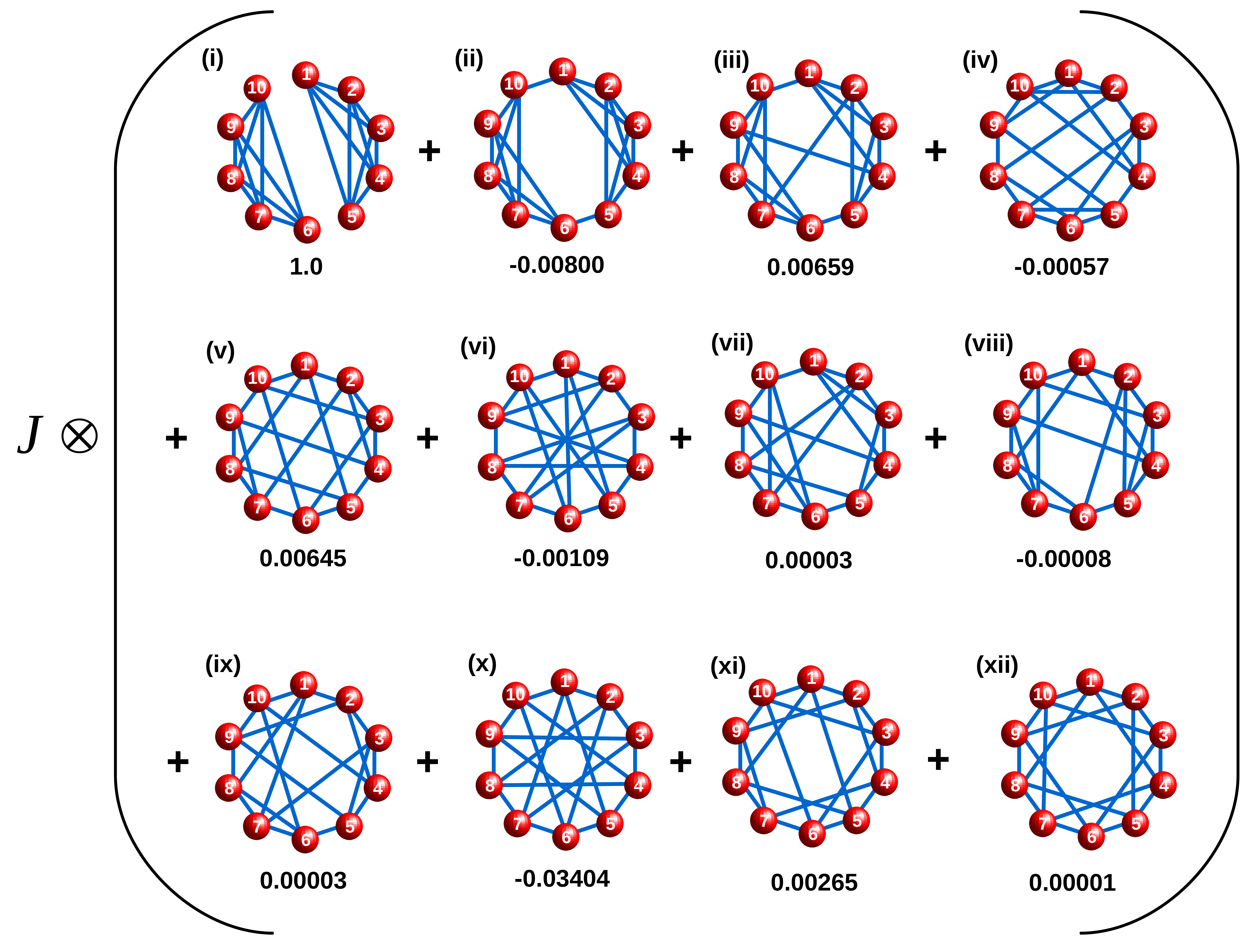}
 	\caption{{\bf Graphs for N=10.} All symbols in the diagram are as described in Fig.\ref{Fig1}. 
 		$J$ represents Jastrow factor for 10 electrons.  Twelve graphs for the set ${\cal S}_{Z2P}^{(10)}=[0,0,0,0,0,2,2,2,2]$ producing linearly independent symmetric functions. The relative weight factors mentioned at the bottom each graph denote the corresponding coefficients of the symmetric functions that provide 0.99999(0) overlap with the exact ground state wave function for the Coulomb interaction.
 	}
 	\label{Fig4}
 \end{figure}

The SOZ for $N=8$ and $10$ respectively are ${\cal S}^{(8)} = [1,1,1,1,3,3,3]$ and ${\cal S}^{(10)} = [1,1,1,1,1,3,3,3,3]$. Decomposing these sets into two distinct sets one of which will reproduce the Jastrow factor $J$ for the respective number of electrons, we obtain other respective sets are ${\cal S}^{(8)}_{Z2P}=[0,0,0,0,2,2,2]$ and ${\cal S}^{(10)}_{Z2P}=[0,0,0,0,0,2,2,2,2]$. One disconnected graph obtainable for each of ${\cal S}^{(8)}_{Z2P}$ and ${\cal S}^{(10)}_{Z2P}$ reproduces Z2P wave function.
Figures \ref{Fig3} and \ref{Fig4} respectively show  graphs that produce linearly independent functions, associated with  ${\cal S}^{(8)}_{Z2P}$ and ${\cal S}^{(10)}_{Z2P}$. Linear combinations of the wave functions that are obtained from the graphs with the corresponding weight factors separately for $N=8$ and $10$ produce exact ground state wave functions  for the Coulomb interaction as the corresponding overlaps are almost 100\% (Table-\ref{Table1}). The number of graphs for every systems of particles producing linearly independent functions are same as the number of states in the zero angular momentum sector when the interaction Hamiltonian is exactly diagonalized. Therefore these graphs cover full Hilbert space in this sector.

Since SOZ up to $N=10$ providing graphs with preserving topological structure generate exact ground states, we believe that SOZs in the thermodynamic limit will also describe the exact ground states. In general, the SOZ for $N$ electrons with Pfaffian shift having $N_\phi = 2N-3$, ${\cal S}^{(N)}$ will have elements $1$ and $3$ with respective numbers $N/2$ and $N/2-1$. Although the actual ground state wave function may differ with interaction, the topological structure is universal; different linear combinations of the microscopic structures (graphs) with the same topological structure determine the ground states for different interactions.

More than $99.999\%$ overlaps (Table-\ref{Table1}) up to $N=10$ have been achieved with appropriate linear combination of the derived linearly independent basis states. The basis states have a common Jastrow factor $J$ and different symmetric functions obtained from the corresponding graph. The symmetric function will not vanish even if the position of some of the particles be same. The maximum number of particles that may coalesce can be determined from the corresponding graph (see methods) by counting the maximum number of vertices that are not connected to each other.  
  Apart from the first one, the graphs for both $N=8$ and 10 in respective figures \ref{Fig3} and \ref{Fig4} contribute even when more than two particles coalesce, in contrary to the disconnected graphs that describe $Z_2$ para-fermion or Pfaffian state. These particles can be viewed as composite bosons as the Jastrow factor $J$ is factored out. 
We identify graphs correspond to the coalescence of maximum possible number of composite bosons. Considering the last graph for both $N=8$ and $N=10$, we find that 4 and 5 composite bosons respectively can coalesce at a point. 
As suggested by the sequence of graphs in Figs.\ref{Fig3} and \ref{Fig4}, up to $N/2$ composite bosons can come close together in the exact state.
This clustering of thermodynamically large number of composite bosons indicates that the Coulomb interaction between electrons in 5/2 state provides the 
strong-coupling regime of the effective interaction between composite bosons. This will even be true in the presence of Landau level mixing or disorder as the topological structure remains invariant. Also, this study indicates that interaction between real fermions may become important for actual neutral edge mode \cite{Wen91} for the enigmatic 5/2 state. Whether or not our wave function is adiabatically connected to the Pf wave function remains to be investigated.

The method presented here for obtaining topological structure and an accurate ground state wave function for Pfaffian flux at 5/2 state, should be applicable to any fractional quantum Hall effect state, and thus is potentially important for revealing the nature of those fractional quantum Hall effect states which are still elusive.
 We believe that the novel topological structure will pave the right path for unraveling the true character of the 5/2 state.

 \vspace{3cm}

 
 {\bf Methods}
 
 Here we outline how the many body wave functions can be obtained from the graphs. As mentioned in the caption of Fig.\ref{Fig1}, the different types of lines joining two electrons describe mutually felt different order of zeros. This can be functionally represented by $z_{ij}^n$ where the exponent $n$ is same as the order of zero felt between $i$-th and $j$-th electrons. Therefore, the central graph in Fig.\ref{Fig1}a produces the function $J^2 = \prod_{i<j}^4 z_{ij}^2$ which is the Jastrow factor for 4 electrons with exponent $2$. The right graph produces the function $z_{12}^{-1}z_{34}^{-1}$ but this function will change if permutatin of the particles are considered. If the links are drawn between 1 and 3 and between 2 and 4, we obtain the function $z_{13}^{-1}z_{24}^{-1}$, and similarly  links between 1 and 4 and between 2 and 3 produce the function $z_{14}^{-1}z_{23}^{-1}$. These three independent functions can be combined to obtain the anti-symmetric wave function
 \begin{equation}   		  	  
 \Psi_{Pf}^{(4)} = \prod_{i<j}^4 z_{ij}^2 \left[ z_{12}^{-1}z_{34}^{-1}-z_{13}^{-1}z_{24}^{-1}+z_{14}^{-1}z_{23}^{-1} \right]		    
 \end{equation}	
 which is nothing but the Pf wave function for 4 electrons upon multiplication of the antisymmetric function obtained from the right graph and the symmetric function $J^2$ obtained from central graph.
 
 The central graph of Fig.\ref{Fig1}b produces the Jastrow factor $J = \prod_{i<j}^4 z_{ij}$ for 4 electrons. The right graph produces the function $z_{12}^2z_{34}^2$. By changing the links between electrons but without altering the set ${\cal S}_{Z2P}^{(4)}$, we obtain two other functions, namely, $z_{13}^2z_{24}^2$ and $z_{14}^2z_{23}^2$. Combining all these functions, we obtain 
 \begin{equation}
 \Psi_{Z2P}^{(4)} = \prod_{i<j}^4 z_{ij} \left[z_{12}^2z_{34}^2+z_{13}^2z_{24}^2+z_{14}^2z_{23}^2   \right]
 \end{equation}
 which is precisely the Z2P wave function for 4 electrons.

 The left graph of Fig.\ref{Fig2}a produces the Jastrow factor $J^2$ for 6 electrons as the links are homogeneous between all the electrons. The right graph of it produces the function $z_{12}^{-1}z_{34}^{-1}z_{56}^{-1}$.  The total number of terms obtained by the all possible permutations are $5!!=15$. Combining all these terms, we obtain the anti-symmetric wave function
 \begin{eqnarray}   		  	  
 \Psi_{Pf}^{(6)} = \prod_{i<j}^6 z_{ij}^2 & & \left[ z_{12}^{-1}\left\{ z_{34}^{-1}z_{56}^{-1} -z_{35}^{-1}z_{46}^{-1}
 + z_{36}^{-1}z_{45}^{-1} \right\}  -	z_{13}^{-1}\left\{ z_{24}^{-1}z_{56}^{-1} -z_{25}^{-1}z_{46}^{-1}
 + z_{26}^{-1}z_{45}^{-1} \right\} \right. \nonumber \\
 & &  + z_{14}^{-1}\left\{ z_{23}^{-1}z_{56}^{-1} +z_{35}^{-1}z_{26}^{-1}
 - z_{36}^{-1}z_{25}^{-1} \right\}  -	z_{15}^{-1}\left\{ -z_{24}^{-1}z_{36}^{-1} +z_{23}^{-1}z_{46}^{-1}
 + z_{26}^{-1}z_{34}^{-1} \right\} \nonumber \\		
 & & \left. + z_{16}^{-1}\left\{ z_{34}^{-1}z_{25}^{-1} -z_{35}^{-1}z_{24}^{-1}
 + z_{23}^{-1}z_{45}^{-1} \right\}\right]    
 \end{eqnarray}
 which is precisely the Pf wave function for 6 electrons. Similarly, the top right graph of Fig.\ref{Fig2}b will generate the function $z_{12}^2z_{26}^2z_{16}^2z_{45}^2z_{34}^2z_{35}^2$. All possible permutation of the particles in different vertices with same topology of the graph will generate $^6C_3/2=10$ independent terms. And the top left graph of Fig.\ref{Fig2}b contributes antisymmetric function $J$. They combine to form the wave function
 \begin{eqnarray}
 \Psi_{Z2P}^{(6)} = \prod_{i<j}^6 z_{ij} & & \left[ z_{12}^2z_{13}^2z_{23}^2 z_{45}^2z_{46}^2z_{56}^2 +z_{12}^2z_{14}^2z_{24}^2z_{35}^2z_{36}^2z_{56}^2 +z_{12}^2z_{15}^2z_{25}^2z_{34}^2z_{36}^2z_{46}^2 +z_{12}^2z_{16}^2z_{26}^2z_{34}^2z_{35}^2z_{45}^2 \right.  \nonumber \\ 
 & & + z_{13}^2z_{14}^2z_{34}^2z_{25}^2z_{26}^2z_{56}^2 + z_{13}^2z_{15}^2z_{35}^2z_{24}^2z_{26}^2z_{46}^2 + z_{13}^2z_{16}^2z_{36}^2z_{24}^2z_{25}^2z_{45}^2   \nonumber \\
 & & +z_{14}^2z_{15}^2z_{45}^2z_{23}^2z_{26}^2z_{36}^2 + z_{14}^2z_{16}^2z_{46}^2z_{23}^2z_{25}^2z_{35}^2 + z_{15}^2z_{16}^2z_{56}^2z_{23}^2z_{24}^2z_{34}^2
 \left. \right]
 \end{eqnarray}
 which is again precisely Z2P wave function for 6 electrons. While the bottom-left diagram of Fig.\ref{Fig2}b again corresponds to the function $J$ for 6 electrons, the bottom-right diagram of it produces the function $z_{12}^2z_{23}^2z_{34}^2z_{45}^2z_{56}^2z_{16}^2$. Without changing the topology of this graph, the permutation of different particles at different vertices will generate $^5C_2\times 3!=60$ independent terms. Considering all these terms, we obtain the corresponding additional wave function as
 \begin{eqnarray}
 & & \Psi_{Ad}^{(6)} = \prod_{i<j}^6 z_{ij} \nonumber \\
 & & \times \left[ z_{12}^2 z_{16}^2 \left\{ z_{23}^2 z_{45}^2 \left( z_{56}^2 z_{34}^2 +z_{46}^2 z_{35}^2\right) + z_{24}^2 z_{35}^2\left( z_{56}^2z_{34}^2 + z_{36}^2 z_{45}^2 \right) +z_{25}^2z_{34}^2 \left( z_{36}^2z_{45}^2 + z_{46}^2z_{35}^2 \right) \right\} \right. \nonumber \\
 & & + z_{12}^2z_{15}^2 \left\{ z_{23}^2z_{46}^2\left(z_{45}^2z_{36}^2+z_{56}^2z_{34}^2 \right) + z_{24}^2z_{36}^2 \left( z_{35}^2z_{46}^2 + z_{56}^2 z_{34}^2 \right) +z_{26}^2z_{34}^2\left( z_{36}^2z_{45}^2 + z_{46}^2z_{35}^2\right)   \right\} \nonumber \\
 & & + z_{12}^2z_{14}^2 \left\{ z_{23}^2z_{56}^2 \left( z_{35}^2z_{46}^2+z_{36}^2z_{45}^2 \right) +z_{25}^2z_{36}^2 \left(z_{35}^2z_{46}^2 +z_{34}^2z_{56}^2 \right) + z_{26}^2z_{35}^2 \left( z_{34}^2z_{56}^2+z_{45}^2z_{36}^2 \right)  \right\} \nonumber \\
 & & +z_{12}^2z_{13}^2 \left\{ z_{24}^2 z_{56}^2 \left( z_{35}^2 z_{46}^2+z_{36}^2z_{45}^2 \right) + z_{25}^2z_{46}^2  \left( z_{45}^2z_{36}^2 +z_{34}^2z_{56}^2   \right) + z_{26}^2z_{45}^2 \left( z_{34}^2z_{56}^2 + z_{35}^2z_{46}^2  \right)  \right\} \nonumber \\
 & & + z_{13}^2z_{16}^2 \left\{ z_{23}^2z_{45}^2 \left( z_{24}^2z_{56}^2 +z_{25}^2z_{46}^2 \right) +z_{34}^2z_{25}^2 \left( z_{24}^2z_{56}^2 +z_{45}^2z_{26}^2   \right) + z_{35}^2z_{24}^2 \left( z_{25}^2z_{46}^2 + z_{45}^2z_{26}^2   \right)   \right\}   \nonumber \\
 & & + z_{13}^2z_{15}^2 \left\{ z_{23}^2z_{46}^2 \left( z_{24}^2z_{56}^2+z_{26}^2z_{45}^2   \right) +z_{34}^2z_{26}^2 \left( z_{24}^2z_{56}^2 +z_{46}^2z_{25}^2   \right)  + z_{36}^2z_{24}^2 \left( z_{26}^2z_{45}^2 + z_{46}^2z_{25}^2   \right)    \right\}  \nonumber \\
 & & + z_{13}^2z_{14}^2 \left\{ z_{23}^2z_{56}^2 \left(z_{25}^2z_{46}^2 + z_{26}^2z_{45}^2   \right)  +z_{35}^2z_{26}^2 \left( z_{25}^2z_{46}^2 + z_{24}^2z_{56}^2  \right) + z_{36}^2z_{25}^2 \left(z_{26}^2z_{45}^2 +z_{56}^2z_{24}^2   \right)    \right\}  \nonumber \\
 & & + z_{14}^2z_{16}^2 \left\{ z_{24}^2z_{35}^2 \left( z_{23}^2z_{56}^2 + z_{25}^2z_{36}^2    \right) + z_{34}^2z_{25}^2 \left( z_{23}^2z_{56}^2 + z_{35}^2z_{26}^2   \right) + z_{45}^2z_{23}^2 \left(z_{25}^2z_{36}^2 + z_{35}^2z_{26}^2  \right)    \right\}  \nonumber \\
 & & + z_{14}^2z_{15}^2 \left\{ z_{24}^2z_{36}^2 \left( z_{23}^2z_{56}^2 + z_{26}^2z_{35}^2   \right) + z_{34}^2z_{26}^2 \left( z_{23}^2z_{56}^2+z_{36}^2z_{25}^2   \right) + z_{46}^2z_{23}^2 \left( z_{26}^2z_{35}^2 + z_{36}^2z_{25}^2  \right)      \right\} \nonumber \\
 & & + z_{15}^2z_{16}^2 \left\{ z_{25}^2z_{34}^2 \left( z_{23}^2z_{46}^2 +z_{24}^2z_{36}^2   \right) + z_{35}^2z_{24}^2 \left( z_{23}^2z_{46}^2 + z_{26}^2z_{34}^2   \right)  + z_{45}^2z_{23}^2 \left( z_{24}^2z_{36}^2 + z_{34}^2z_{26}^2   \right)     \right\}   \left. \right]
 \end{eqnarray}
 We find that the overlap of the state $\Psi^{(6)} = \Psi_{Z2P}^{(6)} + 0.41820 \,\Psi_{Ad}^{(6)}$ with the exact Coulomb ground state, $\frac{\langle \Psi_{ex}^{(6)} \vert \Psi^{(6)}\rangle }{\sqrt{ \langle \Psi_{ex}^{(6)} \vert \Psi_{ex}^{(6)} \rangle
 		\langle  \Psi^{(6)} \vert \Psi^{(6)} \rangle}} \sim 0.99999(0)$ with zero Monte Carlo uncertainty up to five significant digits, where $\Psi_{ex}^{(6)}$ is the ground state obtained by exact diagonalization of the Coulomb Hamiltonian in spherical geometry. 
 
 
  We now show how the symmetric functions obtained from the graphs be nonzero even when some of the particles, {\it i.e.}, the composite bosons meet at the same point. The symmetric function for $N=4$ in Eq.(2),
  $z_{12}^2z_{34}^2+z_{13}^2z_{24}^2+z_{14}^2z_{23}^2$ is nonzero when any two particles meet at a point, say, $z_1=z_3$. The function becomes $2z_{12}^2z_{14}^2$ which will be zero if $z_1 = z_2 $ or $z_1 = z_4$. This suggests that any two particles can meet at a point and other particles will feel zeros of order three (two from the symmetric function and one from $J$) at that point. In general, the maximum number of particles that can be present at a point may be obtained from the corresponding graph by counting the maximum number of vertices that do not have links between themselves, for example the two vertices corresponding to particles 1 and 3 in the right most graph of Fig.1b. If there are $n$ such vertices, $n$ particles can coalesce at a point and the rest of the particles will feel zeros of order $(2^{n-1}+1)$ at that point.
  
  In the similar manner, the symmetric functions in Eq.(4) and Eq.(5) will be nonzero for the coalescence of maximum 2 and 3 numbers of particles respectively for $N=6$. These can also be realized from the respective graphs displayed in Fig.2b. For instance, the vertices correponding to particles 1 and 3 can be the same point in the right-upper panel of Fig2b, while the right-lower pannel suggests vertices 1, 3, and 5 can be the same point.
  
  In Fig.3, the respective graphs support coalescence of 2, 3, 3, and 4 number of particles as for instance vertices corresponding to particles $(1;5)$, $(1;4;6)$, $(1;3;6)$, and $(1;3;5;7)$ respectively are eligible to converge into a point. Similarly, in Fig.4, the respective graphs show the eligibility of convergence of the following vertices: $(1;6)$, $(1;5;7)$, $(1;5;7;9)$, $(1;3;5;8)$, $(1;3;7)$, $(1;3;9)$, $(1;5;7;9)$, $(1;3;6;9)$, $(1;3;6;9)$, $(1;3;6;8)$, $(1;3;7)$, and $(1;3;5;7;9)$ for $N=10$. Therefore, in general, up to $N/2$ particles are allowed to coalesce at a point.       
 

  \vspace{2cm}		  
  		  
 {\bf Acknowledgements}\\
 One of us (SSM) acknowledges support from SRIC, Indian Institute of Technolgy, Kharagpur under grant IIT/SRIC/PHY/EFH/2016-17/178.\\
 
 

 \begin{table}[htb]
 	\caption{{\bf Overlap of the constructed wave function with the exact ground state wave function.}  $\vert \Psi_{ex} \rangle$, $\vert \Psi_{Pf}\rangle$, and $\vert \Psi_{var} \rangle$ respectively represent exact, Pfaffian, and the constructed variational state in Figs.\ref{Fig1}--\ref{Fig4}. $2Q$ represents number of flux quanta for $N$ electrons in spherical geometry. Overlaps ${\cal O}_{Pf} = \frac{\langle \Psi_{ex} \vert \Psi_{Pf}\rangle}{	\sqrt{\langle \Psi_{ex} \vert \Psi_{ex} \rangle \langle  \Psi_{Pf}\vert   \Psi_{Pf} \rangle } } $
 		 is taken from the book \cite{Book}.
 		 Here ${\cal O}_{var} = \frac{	\langle \Psi_{ex} \vert \Psi_{var}\rangle}{	\sqrt{\langle \Psi_{ex} \vert \Psi_{ex} \rangle \langle  \Psi_{var}\vert   \Psi_{var} \rangle } }	$ is calculated using Monte Carlo.} 
 	\label{Table1}
 	\begin{tabular}{ c| c |  c | c}\hline\hline
 		$\,N\,$  &$\,2Q\,$ & ${\cal O}_{{\rm Pf}}$  & $
 		{\cal O}_{{\rm var}}	$ \\
 		\hline\hline
 		4  & 5 & 1.0  &  $1.0$ \\\hline
 		6  & 9 & 0.8162 & $0.99999(0)$ \\
 		\hline
 		8  & 13 & 0.8674 &  $0.99999(0)$ \\
 		\hline
 		10 & 17 & 0.8376 &  $0.99999(0)$ \\
 		
 		\hline   
 		
 		\hline
 		
 	\end{tabular}
 		
\end{table}


		
		
		
	

\end{document}